\documentclass[a4paper,twocolumn,11pt]{quantumarticle}
\pdfoutput=1
\usepackage[utf8]{inputenc}
\usepackage[english]{babel}
\usepackage[T1]{fontenc}
\usepackage{amsmath}
\usepackage{hyperref}
\usepackage{bracket}
\usepackage{tikz}
\usepackage{lipsum}
\usepackage{algorithm}
\usepackage[numbers]{natbib}
\usepackage{algpseudocode} % use this instead of algorithmic

\DeclareUnicodeCharacter{200B}{}
\begin{document}

\title{Universal share-based quantum multi–secret image sharing scheme}

\author{Dipak K. Rabari}
\email{dipakrabari.ec@ddu.ac.in}
\affiliation{Department of Electronics \& Communication Engineering, Dharmsinh Desai University, Nadiad (Gujarat),INDIA }
\orcid{0009-0006-5171-0274}
\author{Yogesh K. Meghrajani}
%\email{latex@quantum-journal.org}
%\homepage{http://quantum-journal.org}
\orcid{0000-0003-0693-1239}
%\thanks{You can use the \texttt{\textbackslash{}email}, \texttt{\textbackslash{}homepage}, and \texttt{\textbackslash{}thanks} commands to add additional information for the preceding \texttt{\textbackslash{}author}. If applicable, this can also be used to indicate that a work has previously been published in conference proceedings.}
\affiliation{Department of Electronics \& Communication Engineering, Dharmsinh Desai University, Nadiad (Gujarat),INDIA }
\author{Laxmi S. Desai}
\affiliation{Department of Mathematics, Dharmsinh Desai University, Nadiad (Gujarat),INDIA }
\orcid{0009-0004-2045-2896}
%\author{Cassandra Granade}
%\affiliation{Microsoft Research, Quantum Architectures and Computation Group, Redmond, WA 98052, USA}
%\author{Johannes Jakob Meyer}
%\affiliation{Dahlem Center for Complex Quantum Systems, Freie Universität Berlin, 14195 Berlin, Germany}
%\orcid{0000-0003-1533-8015}
%\author{Victor V. Albert}
%\affiliation{Institute for Quantum Information and Matter \& Walter Burke Institute for Theoretical Physics, Caltech, Pasadena, CA 91125, USA}
%\orcid{0000-0002-0335-9508}
\maketitle

\begin{abstract}
Image security for information has become increasingly critical as internet become more prevalent due to hacking and unauthorized access. To ensure the security of confidential image data, image encryption using visual cryptography plays a crucial role. To share multiple images using visual cryptography, the company organizer utilizes the concept of a universal or common share. Likewise, quantum computing is an emerging technology that facilitates secure communication. The ability of quantum computers to solve certain mathematical problems efficiently threatens the security of many current encryption algorithms. Hence, to leverage the strengths of quantum computing and visual cryptography, this research introduces a novel universal share-based quantum multi-secret sharing technique for secure image communication. Quantum computing enables the scheme to exhibit high resilience to different eavesdropping threats. Consequently, the proposed method offers robust security solution for sharing confidential images across a range of applications, including enterprise data access and military communications.
\end{abstract}
%\begin{keywords}
%	Visual cryptography, Quantum secret sharing, Multi secret sharing, Secret sharing scheme, Universal share. 
%\end{keywords}
\section{Introduction}
By early 2025, the number of global internet users surpassed 5 billion \cite{Petrosyan2025}, marking a significant rise in the following years. In addition, the growing dependence on Internet-based technologies produces enormous volumes of image data. A significant amount of this image data needs to be protected because it contains sensitive information. To protect the data against intrusion, secret sharing (SS) schemes are implemented. This method includes apportioning a secret among a set of individuals, with this method, a group of individuals are given a secret, with each person holding a portion of the whole secret. A certain number of shares can be integrate to rebuild the secret; an individual share alone holds no value.  Secret sharing is a vital aspect with modern cryptography that is essential to secure data access, storage, and transmission. Visual Cryptography (VC), or the visual secret sharing (VSS) scheme, was presented by Naor and Shamir \cite{naor1994visual} in 1995. Every secret pixel is transformed into a block of \textit{x} sub-pixels, resulting in an image known as share that is x times larger than the original secret. This process generates noise-like share images. All the shares are stacked together to recreate the concealed image. A codebook is used to define the color patterns of the shares based on whether the corresponding secret pixel is black or white. However, traditional SS schemes face several limitations, including the requirement for precise pixel alignment, reliance on a codebook, pixel expansion that distorts the aspect ratio of the original image, and poor contrast in the reconstructed image with stacking of the shares. 
\newline
\par
In 1986, Kafri and Keren \cite{kafri1987encryption} introduced three VSS methods based on a 2-out-of-2 random grid (RG). In these schemes, the shares are stacked to reveal the secret image. Two major advantages of these methods are the elimination of pixel expansion and the removal of codebook requirements. As a result, both the secret and share images preserve their aspect ratio. Nevertheless, the technique has certain limitations, such as weak contrast in the recovered image and sensitivity to pixel alignment. Boolean-based SS mechanisms require minimal computational resources and aim to address these issues. The aforementioned techniques typically allow the sharing of only a single secret image. To enhance the sharing capacity, many researchers have proposed the scheme of multi-secret sharing (MSS) scheme. An MSS technique capable of sharing four secret images across various color formats was proposed by Chen \textit{et al.} \cite{chen2012yet}. Additionally, an MSS scheme with RG using pie-shaped shares was introduced by Lin \textit{et al.} \cite{lin2014distortionless} in 2014. Furthermore, various SS techniques continue to be developed for sharing multiple secret images (\cite{chang2018new}, \cite{hoang2024novel}, \cite{nag2020efficient}, \cite{chattopadhyay2021verifiable}, \cite{kabirirad2018t}, \cite{rabari2025universal} ). 
\par
The universal share (UniShare) based approach \cite{joseph2015random} has been integrated into MSS schemes that utilize common shares. This method incorporates \textit{n} shares along with a UniShare, which serves as a common share. To reconstruct all \textit{n} secret images, a total of \textit{n+1} shares is required. Notably, unlike traditional MSS approaches, retrieving a single secret image does not require all \textit{n+1} shares. The UniShare-based system empowers the chief authority of a firm to enhance secrecy, as the UniShare is necessary for decryption. Meghrajani and Mazumdar \cite{meghrajani2016universal} proposed a UniShare-based MSS system utilizing Boolean operations. Furthermore, a recent scheme \cite{rabari2025universal} introduces a dual-decoding capability in MSS through rotating random grids, where the UniShare is a mandatory component for decryption. While these classical visual cryptographic systems have been effective so far, the emerging threat posed by quantum computing is expected to compromise their security. Consequently, developing cryptographic methods resistant to quantum attacks has become a major research challenge. 
\par
As a result, many researchers attempt to leverage quantum features such as Quantum Key Distribution (QKD) \cite{bennett1984update} to enhance the performance of classical cryptographic schemes by extending them into the quantum environment. In 1999, Hillery  \textit{et al.} \cite{hillery1999quantum} utilized the Greenberger-Horne-Zeilinger (GHZ) state to introduce the Quantum Secret Sharing (QSS) system. Moreover, a threshold QSS system was presented using quantum error-correcting code theory \cite{cleve1999share}. In recent years, several QSS schemes have been developed for sharing a single secret image (\cite{liu2019novel}, \cite{wang2019secret}, \cite{joy2018implementation} ). Additionally, recent research schemes are proposed to illustrate the application of quantum computing. ( \cite{Gustiani2025VQD},\cite{ForemanMasanes2025SeedlessDI}, \cite{Jobst2024MPS} ). There is inadequate literature available on the application of quantum computing (QC) capabilities to VSS schemes involving multiple secrets. In 2025, Ma \textit{et al.} \cite{ma2025quantum} proposed a quantum system-based image secret sharing scheme for hierarchical multi-secret image sharing. In this model, Each participant is unable to access any higher-level secret information, but they can independently compute and obtain a number of lower-level secrets, as the decryption of multi-secret images occurs at different hierarchical levels. 
\subsection{Research contributions}
In this paper, a novel UniShare-based \textit{(n, n+1)} Quantum Visual Multi-Secret Sharing (QVMSS) scheme is presented. Each pixel's color information from the original secret image is encoded into an n-qubit quantum superposition state using the proposed approach. Moreover, the use of a quantum expansion mechanism effectively addresses the resolution degradation problem inherent in classical VSS schemes, ensuring that the recovered image is identical to the original. 
\par
To enhance the security and confidentiality of encrypted image data, the proposed scheme integrates quantum cryptography (QC) with UniShare-based MSS scheme. This fusion leverages the strengths of both frameworks and offers several advantages. First, it enhances security by combining QC with UniShare-based MSS scheme, resulting in a robust encryption framework that significantly improves resistance to unauthorized access. Additionally, it ensures perfect secrecy, as the MSS scheme ensures secrecy through UniShare requirements at the receiver side, while QC prevents manipulation and eavesdropping, thereby defending against brute-force and other conventional attacks. Furthermore, the proposed hybrid technique is interoperable with existing SS systems, as it enhances security without necessitating substantial modifications to current quantum communication infrastructures, enabling seamless integration into existing image encryption frameworks. 
\par
To achieve enhanced protection, absolute privacy, and interoperability with existing SS schemes, the proposed approach integrates the concepts of UniShare-based MSS scheme and QC. Specifically, this work presents a novel hybrid SS scheme, grounded in a secure VC framework that combines quantum computing with the UniShare-based MSS technique. By leveraging the computational capabilities of quantum systems alongside the structural efficiency of the MSS scheme, the proposed method pursues an excessive security level for image transmission over untrusted communication channels. Additionally, ensuring the secure transfer of encrypted images is an inherent features of quantum technology. Thus, the proposed scheme introduces a novel UniShare-based MSS approach based on quantum computing for secure image transmission over untrusted and potentially malicious infrastructures. Furthermore, a comparative analysis of recent QVMSS schemes developed for effective image security is also presented. 
\subsection{Paper outline}
The subsequent sections of the paper are organized in this manner. Section II provides a brief overview of the fundamentals of quantum computing. Section III reviews related research on QSS schemes. Section IV describes the proposed \textit{(n, n+1)} QVMSS scheme, including encoding and decoding processes. Section V presents the experimental results using a \textit{(2, 2)} QVMSS scheme and discusses comparisons with other related schemes. Finally, the conclusion is presented in Section VI.
\section{Preliminaries}
With the rapid advancements in quantum computing, a growing number of users are recognizing its potential benefits. Several quantum algorithms, most notably those developed by Grover and Shor, have been developed, posing significant threats to classical cryptographic systems by rendering many conventional encryption methods vulnerable. As a result, classical cryptography (CC) schemes are increasingly viewed as insufficient to meet modern information security requirements. 
\newline
Quantum encryption, a subfield of quantum information science, represents one of the practical applications of quantum computing. It is a physics-based approach that employs the use of the concepts of quantum mechanics to enhance security of data. This method takes advantage of the inherent unpredictability at the quantum level, particularly the behavior of photons, to perform secure encryption and decryption. In contrast to classical and post quantum cryptographic (PQC) methods, which encode information using binary bits, QC makes use of quantum bits, or qubits. 
\newline
Both CC and QC aim to ensure data confidentiality, integrity, and authenticity; however, their underlying mechanisms differ substantially. Classical cryptographic systems are based on the computational intractability of mathematical problems such as large integer factorization and discrete logarithms, which are assumed to be infeasible to solve using conventional computing methods. In contrast, QC relies on the physical properties of quantum mechanics, including superposition and entanglement, for securing data. 
\newline
Interestingly, PQC follows the CC paradigm by employing hard mathematical problems presumed to remain intractable even for quantum computers, thus providing quantum-resistant security. Among the most prominent applications of QC is QKD, which facilitates secure and theoretically unbreakable communication channels. Unlike CC and PQC, QKD can detect eavesdropping attempts by leveraging the no-cloning theorem and related quantum principles, which ensure that any unauthorized measurement of quantum data introduces detectable disturbances in the system. 
\newline
Despite its theoretical robustness, the deployment of QKD requires specialized infrastructure, such as photon emitters and quantum-compatible optical fibers, leading to high implementation costs that can reach several million dollars at enterprise scales. The following section presents the fundamental aspects of quantum computing pertinent to the proposed study.
\subsection{Qubit representation}
The fundamental unit of conventional computing is the bit. An analogous concept, the qubit, forms the foundation of quantum computing. Qubits are typically regarded as conceptual mathematical entities. In quantum computing, the focus is on the quantum state. In Hilbert space, a vector can be used to represent a quantum state. The following is a vector in Hilbert space: 
\begin{equation}
	\ket{\Psi}= \begin{bmatrix} 	0  \\
		1  		\end{bmatrix}
\end{equation}
The Dirac symbol, which is specifically used in quantum mechanics, is represented by the sign $|.\rangle$. It indicates that an object is a vector in Hilbert space. ${\Psi}$ is the label of the vector $>$ is called ket and the corresponding bra is $\bra{\Psi}$,which is the conjugate transpose of the ket. In classical information theory, the bit is the basic unit of information, which can exist in either of the two states 0 or 1. Analogously, in quantum information theory, the fundamental unit is the quantum bit or qubit \cite{wang2022review}. A qubit has two possible basis states, $\ket{0}$ and  $\ket{1}$, and can exist in a quantum superposition of these states, expressed as  $\ket{q}$ = $\alpha$  $\ket{0}$ + $\beta$  $\ket{1}$, where $\alpha$, $\beta$ $\in$ C are complex coefficients that satisfy the normalization condition $|\alpha|^2$ + $|\beta|^2$ = 1. Following the quantum measurement, a particular quantum state is reached by $\ket{q}$. It is likely that it will collapse into $\ket{0}$ with a probability of $|a|^2$ and into $\ket{1}$ with a probability of $|b|^2$. Below is a vector representation of $\ket{0}$  and $\ket{1}$ :
\begin{align}
	\ket{0} &= \begin{bmatrix} 1  \\ 0  \end{bmatrix} & \ket{1} &= \begin{bmatrix} 0  \\ 1  \end{bmatrix}
\end{align}
\subsection{Quantum Gate}
\begin{figure}[t]
	\centering
	\includegraphics[width=0.5\textwidth]{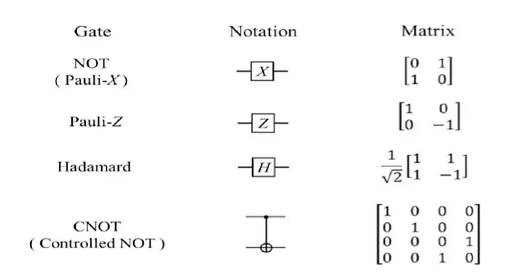}
	\caption{Quantum gate information for Pauli-x, Pauli-Z, Hadamard and CNOT gate}
	\label{fig_1}
\end{figure}

A quantum computer is built using a quantum circuit composed of wires and elementary quantum gates that transport and manipulate quantum information, analogous to how a classical computer is implemented using electrical circuits with wires and logic gates. Among these gates, the Hadamard gate exhibits uniquely quantum behavior by transforming a definite quantum state into a superposition of states, unlike the Pauli gates which do not inherently produce superposition. For example, it can convert a spin-down state into a state that simultaneously represents both spin-up and spin-down. The most basic category of quantum gates includes single-qubit gates such as Pauli-X, Pauli-Z, and Hadamard (H), which are represented by 2$\times$2 unitary matrices, as shown in Figure \ref{fig_1}.
Another essential component of a quantum circuit is the multi-qubit gate. The controlled-NOT (CNOT) gate, illustrated in Figure \ref{fig_1}, is a fundamental two-qubit quantum logic gate that operates on a control qubit and a target qubit. The target qubit remains unaltered if the control qubit is set to 0. Conversely, the target qubit flips if the control qubit is set to 1.
\section{Related work}
The rapidly evolving field of quantum computing has attracted significant attention due to its high potential for securing digital information through quantum communication protocols. To enable secure communication, various quantum computing-based schemes have been proposed by researchers ( \cite{singh2025advancements}, \cite{satre2025quantum}, \cite{prajapat2025quantum} ). Despite notable advancements, several challenges remain unresolved. This section reviews some of the most relevant studies in the field of QC as applied to the security of secret images.

The BB84 protocol, originally proposed by Bennett and Brassard in 1984, has become one of the widely recognized and researched quantum communication techniques \cite{bennett1984update}. Numerous researchers have examined and improved the security and effectiveness of BB84 since its inception, resulting in a number of modifications and enhancements. More recently, quantum computing has been increasingly applied to secret image sharing (SIS) schemes across various domains ( \cite{zhang2025t}, \cite{wang2024t}, \cite{di2025security} ). These schemes are particularly effective for image security, and the integration of QC techniques has been shown to significantly enhance their robustness.

\cite{zhang2025t} presented a \textit{(t, n)} threshold quantum secret sharing (QSS) scheme, wherein any \textit{t} out of \textit{n} users can recover the original secret image. The scheme offers low computational complexity and supports reusability of the identity key. In 2024, Wang \textit{et al.} \cite{wang2024t} presented a technique in which a quantum matrix multiplier is initially employed for encryption. Quantum gates are utilized to generate sample matrices at the transmitter and to collect shadow images at the receiver for reconstruction. Additionally, quantum circuits are designed to perform qubit operations for both encoding and decoding processes. The inherent security features of quantum communication further contribute to the promising performance of QC-based techniques for secure image sharing.

In \cite{di2025security}, the authors reported a QSS scheme based on a generic distributed quantum network utilizing a threshold system. Within this framework, all participants collaboratively route quantum information among themselves. Moreover, the share images in VSS are noisy and meaningless. Consequently, the shared image becomes vulnerable to attacks following quantum measurement. To address this issue, \cite{zhao2023quantum} proposed an \textit{(n, n)} quantum meaningful visual SS scheme aimed at enhancing the security of image data transmission. To ascertain the privacy and authenticity of images transmitted over unsecured communication channels, extensive research has been conducted in the domain of quantum image cryptography. Several of the latest quantum-based image encryption techniques are discussed in ( \cite{joy2018implementation}, \cite{wang2022review}, \cite{satre2025quantum}, \cite{kadhim2024quantum} ).

On the other hand, limited research has been reported on UniShare-based MSS techniques. Furthermore, multi-image secure transmission has typically been achieved using conventional cryptographic methods. However, due to the increasing power of modern computing systems, traditional encryption techniques are becoming insufficient to resist sophisticated attacks. These methods are increasingly vulnerable, particularly given the enhanced computational capabilities of contemporary systems. Therefore, the development of new and more robust cryptographic algorithms is essential, especially for MSS applications.

The motivation for proposing a novel UniShare-based QSS scheme for multi-secret image sharing arises from the limited studies in this area and the suboptimal security performance of current approaches based on QC. This research addresses the critical need for secure and efficient transmission of multiple secret images over unreliable networks. Moreover, as MSS applications become increasingly prevalent in sectors such as organizational data access, defense, and finance, it is imperative to ensure that multiple image data can be transmitted securely within the available channel bandwidth.

The proposed methodology utilizes the principles of quantum physics and the image secret sharing concept from VC to develop a novel cryptographic method that ensures the security of UniShare-based multi-secret image data. A quantum technology communication protocol is employed to securely transmit the qubits generated by the proposed method. The integration of these two techniques provides strong protection against hacking and eavesdropping threats.

\section{Proposed scheme}
The proposed scheme presents a universal-share-based QSS scheme for the secure transmission of multiple secret images. In this technique, a novel RG-based quantum encryption and decryption mechanism is introduced for processing multiple secret images. The proposed QSS scheme is designed to provide an excessive level of secrecy and confidentiality when transmitting multiple secret images over unsecured communication channels. A quantum circuit is developed to process each qubit corresponding to the pixel values of binary images, analogous to classical image processing. In the case of binary secret images, each pixel has a value of either 0 or 1, which is mapped to the quantum states $\ket{0}$ or $\ket{1}$, respectively, for processing within the proposed quantum circuit.

\begin{figure}[!t]
	\centering
	\includegraphics[width=0.5\textwidth]{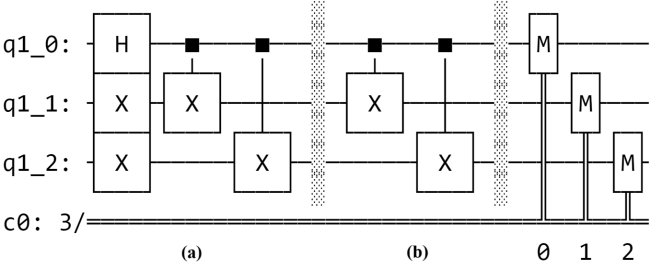}
	\caption{Quantum circuit for UniShare-based MSS (a) Transmitter end (b) Receiver end }
	\label{fig_2}
\end{figure}

The proposed encryption algorithm consists of two stages: the generation of random UniShare and the generation of n random-like shares for \textit{n} multiple secret images. Initially, the declared qubits ($q_0$, $q_1$ and $q_2$) are assumed to be in the $\ket{0}$ state. A Hadamard gate is employed to generate the UniShare, which ensures the random generation of qubit $q_0$​ in the$\ket{0}$  or $\ket{1}$ state within the proposed quantum circuit, as illustrated in Figure \ref{fig_2}. A NOT gate is applied to transform the pixel values of the multiple secret images into quantum states for qubits $q_1$​ and $q_2$​. Furthermore, an RG-based technique \cite{kafri1987encryption} is adopted, which performs pixel-level operations between the UniShare and the \textit{n} secret images. Each pixel in the resulting \textit{n} share images is obtained by performing an XOR operation between the corresponding pixel of the UniShare and each pixel of the \textit{n} secret images. 

Accordingly, the CNOT gate is utilized in the proposed quantum circuit to generate the \textit{n} share images. In this setup, the inputs to the CNOT gate are the qubit values representing the multiple secret images ( $q_1$ and $q_2$ ), while the control qubit is derived from the Hadamard-generated qubit, conceptually referred to as the UniShare. Each output qubit from the CNOT gate is interpreted as a pixel value in the corresponding share image. The pseudocode for generating the encrypted share $S_1$​, using the qubit value of the UniShare U and two secret images $G_1$​ and$G_2$​, demonstrates the operation of the CNOT gate within the encryption algorithm. Each output qubit value of CNOT gate is conceptually considered as each pixel value of all the \textit{n} share images. The Pseudo code for the generation of encrypted share $S_1$ with the qubit value of UniShare U and two secret images $G_1$​ and $G_2$ with the CNOT gate operation are mentioned in encryption algorithm.  
	\begin{algorithm}[H]
		\caption{Encryption algorithm}
		\label{alg:alg1}
		\begin{algorithmic}[1]
			\State \textbf{Input:} Multiple secret images $G_1$, $G_2$
			\State \textbf{Output:} Two encrypted shares $S_1$, $S_2$ and a UniShare $U$
			
			\For{$i = 1$ to $M$}
			\For{$j = 1$ to $N$}
			\State Generate random qubit $q_0 \in \{\ket{0}, \ket{1}\}$ and assign to $U_{ij}$
			
			\Comment{First secret image $G_1$}
			\If{$G_{1ij} = \ket{0}$ and $U_{ij} = \ket{1}$}
			\State $S_{1ij} \gets \ket{1}$
			\ElsIf{$G_{1ij} = \ket{1}$ and $U_{ij} = \ket{0}$}
			\State $S_{1ij} \gets \ket{1}$
			\Else
			\State $S_{1ij} \gets \ket{0}$
			\EndIf
			
			\Comment{Second secret image $G_2$}
			\If{$G_{2ij} = \ket{0}$ and $U_{ij} = \ket{1}$}
			\State $S_{2ij} \gets \ket{1}$
			\ElsIf{$G_{2ij} = \ket{1}$ and $U_{ij} = \ket{0}$}
			\State $S_{2ij} \gets \ket{1}$
			\Else
			\State $S_{2ij} \gets \ket{0}$
			\EndIf
		
			\EndFor
			\EndFor
		\end{algorithmic}
	\end{algorithm}
The decryption phase employs a CNOT gate controlled by qubit $q_0$ (the UniShare), as depicted in Figure \ref{fig_2}. This operation reconstructs the original n multiple secret images. The measurement operation at the output of each CNOT gate represents the quantum state of each measured qubit at the receiver's end. Upon converting the measured qubit values into their binary pixel equivalents, the original multiple secret images are recovered without any loss. The pseudocode outlining the lossless reconstruction of the original secret images at the receiver is presented in the decryption algorithm, which demonstrates the operation of the CNOT gate.
\begin{algorithm}[H]
	\caption{Decryption algorithm}
	\label{alg:alg2}
	\begin{algorithmic}[1]
		\State \textbf{Input:} Two encrypted shares $S_1$, $S_2$ and a UniShare $U$
		\State \textbf{Output:} Reconstructed secret images $G_1'$, $G_2'$
		
		\For{$i = 1$ to $M$}
		\For{$j = 1$ to $N$}
		
		\Comment{First secret image $G_1'$}
		\If{$S_{1ij} = \ket{0}$ and $U_{ij} = \ket{1}$}
		\State $G_{1'ij} \gets \ket{1}$
		\ElsIf{$S_{1ij} = \ket{1}$ and $U_{ij} = \ket{0}$}
		\State $G_{1'ij} \gets \ket{1}$
		\Else
		\State $G_{1'ij} \gets \ket{0}$
		\EndIf
		
		\Comment{Second secret image $G_2'$}
		\If{$S_{2ij} = \ket{0}$ and $U_{ij} = \ket{1}$}
		\State $G_{2'ij} \gets \ket{1}$
		\ElsIf{$S_{2ij} = \ket{1}$ and $U_{ij} = \ket{0}$}
		\State $G_{2'ij} \gets \ket{1}$
		\Else
		\State $G_{2'ij} \gets \ket{0}$
		\EndIf
		
		\EndFor
		\EndFor
	\end{algorithmic}
\end{algorithm}

\section{Experimental results and discussion}
This section presents the simulation results of the proposed scheme. Experiments were conducted to evaluate its effectiveness using 512$\times$512 binary images. The implementation was carried out using Qiskit on Python 3.12, executed on a system with an 11th Gen Intel(R) Core(TM) i3-1115G4 \@ 3.00 GHz CPU and 8 GB of RAM. The proposed scheme is implemented for sharing two secret images, and the corresponding quantum circuit is designed as shown in Figure \ref{fig_2}.
\begin{figure}[!t]
	\centering
	\includegraphics[width=0.5\textwidth]{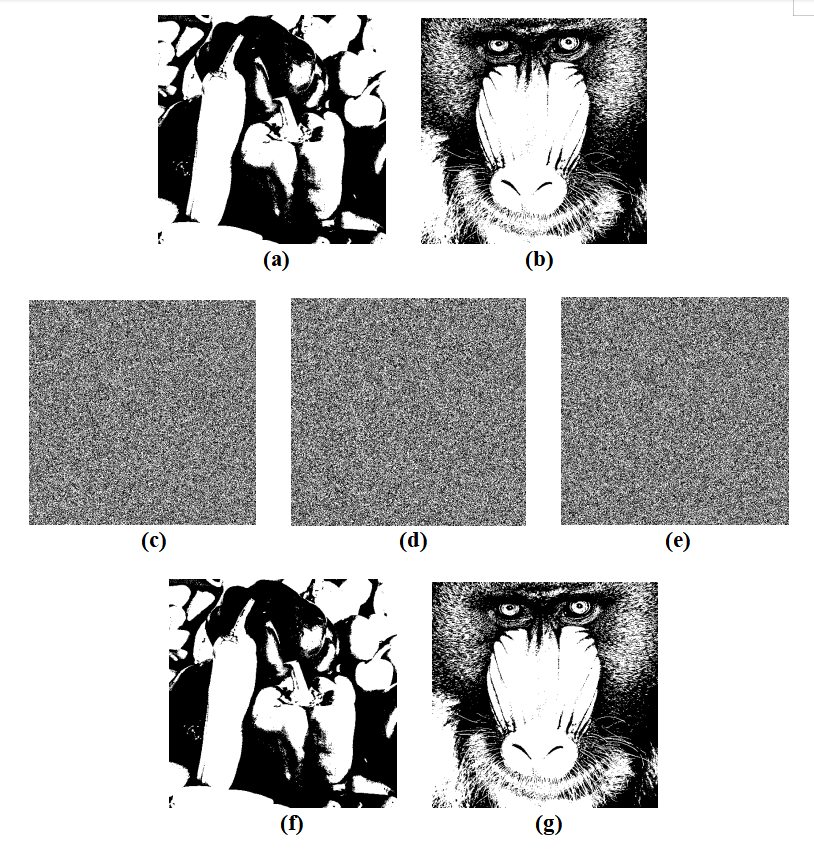}
	\caption{(a) Original Secret Image 1 (b) Original Secret Image 2 (c) UniShare image (d) Encrypted Share Image 1 (e) Encrypted Share Image 2 (f) Recovered Secret Image 1 (g) Recovered Secret Image 2 }
	\label{fig_3}
\end{figure}

The binary secret images consist of pixels that are either transparent or opaque, representing classical bit values of 0 or 1, respectively. Experimental results for the sharing of two binary secret images are illustrated in Figure \ref{fig_3}. The initial states of the qubits are set to either $\ket{0}$ or $\ket{1}$, which are equivalent to classical bit values of 0 or 1 in the binary secret images. Hadamard gates are used to generate random qubit states $\ket{0}$ or $\ket{1}$, which collectively form a random-like UniShare, as depicted in Figure \ref{fig_3}(c). Each qubit corresponding to the secret images in Figure \ref{fig_3}(a) and Figure \ref{fig_3}(b) is applied as input to the CNOT gate. According to the RG algorithm, each encrypted share is generated through a pixel-wise XOR operation between the UniShare and the respective secret images. Accordingly, the output of the CNOT gate produces the encrypted qubit states, where the output of the Hadamard gate serves as the control qubit for the CNOT operation, as shown in Figure \ref{fig_2}. The resulting noise-like encrypted shares are illustrated in Figure \ref{fig_3}(d) and Figure \ref{fig_3}(e). 

At the receiver end, the measurement of these qubit states results in the recovery of the decrypted multi-secret images. The measured outputs of the CNOT gates are converted to classical bits (0 or 1), which represent the pixel values of the reconstructed secret images, as shown in Figure \ref{fig_3}(f) and Figure \ref{fig_3}(g). The proposed approach successfully achieves lossless recovery of the multiple secret images. A key advantage of this scheme lies in the enhanced security of the shares, as any attempt by an intruder to measure the qubits of the encrypted shares will fail due to quantum uncertainty.

\subsection{Performance analysis}
In this section, the proposed scheme is assessed with the quality parameters such as PSNR, SSIM, and correlation as mentioned below:
\subsubsection{Peak Signal Noise Ratio \cite{rabari2024lock}}
An objective assessment of an image's quality based on the difference in pixel values between two images is the Peak Signal Noise Ratio (PSNR). A higher PSNR value means that the restored image has better visual quality. The mathematical expression for PSNR is represented by Equation \eqref{psnr}.
\begin{equation}
	\label{psnr}
	PSNR=10 * \log_{10}\frac{MAX^2}{MSE} dB
\end{equation}
Where MAX=255 for grayscale image and MSE is mean square error which is expressed as shown in Equation \eqref{mse}
\begin{equation}
	\label{mse}
	MSE=\frac{1}{mn} \sum\limits_{i=1}^m \sum\limits_{j=1}^n (A_{ij}-B_{ij})^2
\end{equation}
m      = number of rows of an image\\
n      = number of columns of an image\\
$A_{ij}$ = pixel value from original image\\
$B_{ij}$ = pixel value from recovered image\\
\begin{figure}[!t]
	\centering
\includegraphics[width=0.5\textwidth]{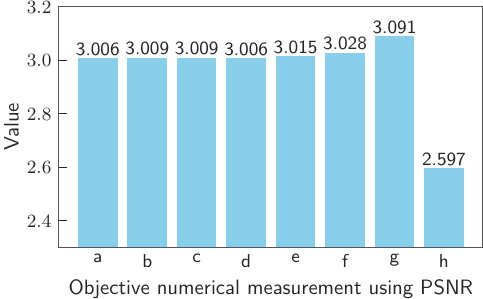}
	\caption{Objective numerical measurement using PSNR (a) Secret image 1 Vs share image 1 (b) Secret image 1 Vs share image 2 (c) Secret image 2 Vs share image 1 (d) Secret image 2 Vs share image 2 (e) Secret image 1 Vs UniShare image  (f) Secret image 2 Vs UniShare image (g) Share image 1 Vs UniShare image (h) Share image 2 Vs UniShare image }
	\label{fig_4}
\end{figure}
\subsubsection{Structural Similarity Index Measure \cite{rabari2024lock}}
Structural Similarity Index Measure (SSIM) is utilized to estimate the similarity between two images. Higher SSIM value implies better similarity between two images. For two identical images, SSIM value is 1. It is expressed mathematically as shown in Equation \eqref{SSIM}.
\begin{equation}
	\label{SSIM}
	SSIM(x,y) = \frac{(2\mu_x\mu_y + C_1) + (2 \sigma _{xy} + C_2)} 
	{(\mu_x^2 + \mu_y^2+C_1) (\sigma_x^2 + \sigma_y^2+C_2)}
	%	\label{eq:SSMI}
\end{equation} 
Where $\mu_x$ = mean for x, $\mu_y$ = mean for y, $\sigma_x^2$ = the variance of x, $\sigma_y^2$ = variance of y, $\sigma _{xy}$ = covariance of x and y, $C_1$= $(k_1*L)^2)$  and $C_2$= $(k_2*L)^2)$ are two variables to stabilize the division with weak denominator, L is the dynamic range of the pixel-values, $k_1$=0.01 and $k_2$=0.03 by default.
\subsubsection{Correlation \cite{rabari2025universal} } 
Correlation is a measure to express interrelation between two images. Higher value of correlation represents closeness of recovered image with original secret image. Correlation is expressed as shown in Equation \eqref{cor},
\begin{equation}
	\label{cor}
	\rho(x,y)={\frac{\textit{cov (x,y)}}{\sigma_x\sigma_y}}
\end{equation}
\begin{figure}[!t]
	\centering
\includegraphics[width=0.5\textwidth]{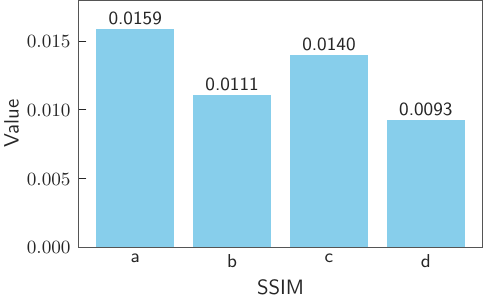}
	\caption{Objective numerical measurement (a) SSIM between secret image 1 Vs share image 1 (b) SSIM between secret image 2 Vs share image 2 (c) Correlation between secret image 1 Vs share image 1 (d) Correlation between secret image 2 Vs share image 2  }
	\label{fig_5}
\end{figure}
The low PSNR values of the secret images and corresponding share images, as shown in Figure \ref{fig_4}, indicate that no confidential information is leaked if an adversary gains access to the share images alone. The UniShare, in conjunction with the share images, is essential for the successful reconstruction of the multiple secret images at the receiver. Furthermore, Figure \ref{fig_5} represents the quality analysis of the presented scheme, including the SSIM and correlation coefficients calculated between the secret images and the share images. The secret images and the reconstructed images at the receiver exhibit a high degree of structural similarity and correlation, both achieving values of 1. In contrast, the low SSIM and correlation values between the secret images and the share images demonstrate the absence of structural similarity or statistical dependence, thereby reinforcing the confidentiality and security offered by the proposed method.

\subsection{Comparison of QSS schemes}
The comparison between the proposed method and existing MSS schemes is presented in Table \ref{tab:Table1}. The proposed technique offers MSS capabilities compared to Liu \textit{et al.} \cite{liu2019novel} and Wang \textit{et al.} \cite{wang2024t}, making it suitable for a wider range of secret sharing applications. Moreover, the proposed scheme enables lossless recovery of multiple secret images at the receiver. In terms of security, the suggested method provides enhanced protection due to the essential requirement of the UniShare at the receiver, unlike the \textit{(t,n)} MSS scheme in \cite{ma2025quantum}. Since the UniShare is held exclusively by a principal authority, the scheme is resilient against attacks involving dishonest participants. Even if an adversary gains access to all other shares, the secret images cannot be reconstructed without the UniShare. Compared to the approaches in ( \cite{hoang2024novel}, \cite{nag2020efficient}, \cite{chattopadhyay2021verifiable}, \cite{kabirirad2018t}, \cite{rabari2025universal} ), the proposed scheme leverages the advantages of quantum computing in addition to supporting MSS functionalities. Unlike \cite{liu2019novel} and \cite{samadder2022quantum}, the proposed method follows an \textit{(n, n+1)} structure by utilizing a UniShare. Furthermore, our scheme eliminates the need for chaotic image generation when producing \textit{n+1} shares, as required in \cite{luo2019novel}. Notably, in the proposed method, lossless recovery of the n original secret images can be obtained using the UniShare along with any one of the n encrypted shares. This is in contrast to \cite{luo2019novel}, which requires all \textit{n+1} shares for successful reconstruction at the receiver.
\begin{table}[!t]
	\centering
	\caption{Comparative analysis of recent QSS Schemes } 
	% \newline
	\label{tab:Table1}
	\centering
	\resizebox{\columnwidth}{!}{
	\begin{tabular}{|c|c|c|c|c|c|}
		\hline
		Authors & No. of& Image & Type & Shares  & Uni-\\
		& secrets & format & of   &  required & Share\\
		& &  &  VSS & for  & \\
		& &  &   &  recovery & \\
		\hline
		Liu &Single&Binary &\textit{(n,n)}&\textit{n}&No\\
		\textit{et al.} \cite{liu2019novel}&&&&shares&\\
		(2019)&&&&&\\
		\hline
		Luo   &Multi&Grayscale &\textit{(n,n+1)}&\textit{n+1}&No\\
		\textit{et al.} \cite{luo2019novel}&&&&shares&\\
		(2019)&&&&&\\
		\hline
		Liu 	 &Single&Binary & \textit{(t,n)}&\textit{t}&No\\
		\textit{et al.} \cite{liu2020t}&&&&shares&\\
		(2020)&&&&&\\
		\hline
		Wang   &Single	&Grayscale & \textit{(t,n)}&\textit{t}&No\\
		\textit{et al.} \cite{wang2024t}&&and color&&shares&	\\
		(2024)&&&&&\\
		\hline
		Chaudhry 	 &Multi	&Binary & \textit{(n,n)}&\textit{n}&No\\
		%			Secret Image3 &&&&	\\
		\& Dutta \cite{samadder2022quantum} &&&&shares&	\\
		(2022)&&&&&\\
		\hline
		Ma    &Multi	&Binary &\textit{(t,n)} &\textit{t}&No\\
		%			Secret Image4 &&&&	\\
		\textit{et al.} \cite{ma2025quantum}&&&&shares&\\
		(2025)&&&&&\\
		\hline
		Ours	& Multi& Binary & \textit{(n,n+1)}&\textit{n+1}&Yes\\
		(2025)&&&&&\\
		\hline
	\end{tabular}}
\end{table}
\section{Conclusion}
In this study, a novel QSS scheme for securing multiple images has been proposed. The scheme integrates UniShare-based MSS with quantum technology to ensure the security and privacy of information about secret images sent over communication channels. Experimental results demonstrate that the proposed model achieves strong performance in terms of both security and encryption efficiency. The inherent properties of the quantum system make it highly resistant to interception and decoding by unauthorized parties lacking access to the UniShare. Furthermore, the proposed method exhibits robustness against the attacks.

\bibliographystyle{quantum}
\bibliography{reference}

\end{document}